%% file: main.tex
\documentclass[journal]{IEEEtran}
\usepackage{graphicx}
\usepackage{amsmath,amssymb,amsfonts,dsfont}
\usepackage{stmaryrd}
\usepackage{url}
\usepackage{dsfont}
\usepackage{float}
\usepackage{hyperref}
\usepackage{multirow}

\include{notations}
\usepackage{adjustbox}
\usepackage{graphicx}
\usepackage{siunitx}
\usepackage{hhline}
\usepackage[dvipsnames,table,xcdraw]{xcolor}
\usepackage[caption=false]{subfig}

\ifCLASSINFOpdf
\else
\fi

\begin{document}
%
\title{Dreem Open Datasets: Multi-Scored Sleep Datasets to compare Human and Automated sleep staging}
%
%
%

\author{Antoine~Guillot, Fabien~Sauvet, Emmanuel~H~During and~Valentin~Thorey
\thanks{A. Guillot and V. Thorey are with the Algorithm Team, Dreem, Paris

E. H. During is with the Center for Sleep Sciences and Medicine, Stanford University, Stanford, California, USA

F. Sauvet is with the French Armed Forces Biomedical Research Institute (IRBA), Fatigue and Vigilance Unit, Bretigny sur Orge, France; EA 7330 VIFASOM, Paris Descartes University, Paris, France
}}

\maketitle

\input{0_abstract.tex}


%
\IEEEpeerreviewmaketitle

\input{1_introduction.tex}

\input{2_materials_and_methods.tex}

\input{3_experiments.tex}

\input{4_discussion.tex}

\input{5_conclusion.tex}


%

\section*{Acknowledgment}

We would like to thank the Fatigue and Vigilance team including Drogou C., Erblang M., Dorey R., Quiquempoix M., Gomez-Merino D. and Rabat A. for their help in the clinical trial at the Institut de Recherche Biologique des Arm\'ees (IRBA).  We would like to thanks Dr. Michael E. Ballard, Hugo Jourde, Polina Davidenko, Sarah deLanda and Stephanie  Lettieri for  their help  in  realizing  the  clinical  trial  at  the  Stanford  Sleep Medicine Center. We also would like to thank the Dreem team, including Mason Harris for his help gathering both datasets together with the sleep technologists' scorings, and Pierrick Arnal, Th\'eo Moutakanni, Cl\'emence Pinaud and Olivier Tranzer for their help with the manuscript.

\bibliographystyle{IEEEtran}
\bibliography{IEEEabrv,references.bib}

\ifCLASSOPTIONcaptionsoff
  \newpage
\fi
\clearpage
\pagebreak
\appendices



%

\end{document}

%% file: 0_abstract.tex
\begin{abstract}
Sleep stage classification constitutes an important element of sleep disorder diagnosis. It relies on the visual inspection of polysomnography records by trained sleep technologists. Automated approaches have been designed to alleviate this resource-intensive task. However, such approaches are usually compared to a single human scorer annotation despite an inter-rater agreement of about 85 \% only. The present study introduces two publicly-available datasets, DOD-H including 25 healthy volunteers and DOD-O including 55 patients suffering from obstructive sleep apnea (OSA). Both datasets have been scored by 5 sleep technologists from different sleep centers. We developed a framework to compare automated approaches to a consensus of multiple human scorers. Using this framework, we benchmarked and compared the main literature approaches to a new deep learning method, SimpleSleepNet, which reach state-of-the-art performances while being more lightweight. We demonstrated that many methods can reach human-level performance on both datasets. SimpleSleepNet achieved an F1 of 89.9 \% vs 86.8 \% on average for human scorers on DOD-H, and an F1 of 88.3 \% vs 84.8 \%  on DOD-O. Our study highlights that state-of-the-art automated sleep staging outperforms human scorers performance for healthy volunteers and patients suffering from OSA. Considerations could be made to use automated approaches in the clinical setting.

\end{abstract}

%% file: 1_introduction.tex
\section{Introduction}
%

\IEEEPARstart{S}{leep} has a crucial impact in human health. Sleep disorders are a common public health issue. For instance, in the US, studies have shown that millions of people are affected  \cite{NationalCenteronSleepDisordersResearchandothers2011NationalPlan}. Polysomnography (PSG) is the gold standard for the diagnosis of highly prevalent sleep disorders such as obstructive sleep apnea (OSA). It consists of recording various bio-physiological signals such as electroencephalogram (EEG), electrooculogram (EOG), electromyogram (EMG), and can include breathing and cardiac signals. Sleep stage classification consists of the visual inspection and classification of 30-seconds epochs of PSG by sleep technologist. The output of this process is the hypnogram, the diagram of sleep stages throughout the night. It is a systematic and valuable preliminary step in performing a diagnosis. Sleep stages are labeled by sleep technologist following the American Association of Sleep Medicine (AASM) rules \cite{Iber2007TheSpecifications}. These rules set out 5 stages, based on the various waveforms observed on each signal of the PSG: wake, rapid eye movement (REM), non-REM sleep stage 1 (N1), 2 (N2) and 3 (N3). It typically takes a sleep technologist 30 minutes to an hour to perform sleep staging on a whole record, i.e. about one thousand 30-second epochs, making it time-consuming and expensive. Another important aspect of sleep staging is the relatively low inter-rater agreement. Indeed, by definition, the AASM rules act as guidelines but do not fully characterize the natural variability that a PSG signal can measure. Hence, a study conducted on the AASM Inter-scorer Reliability dataset shows an average inter-rater agreement of 82.6\% using sleep stages from more than 2,500 experimented sleep scorers \cite{Rosenberg2014TheEvents}. Agreement varies between sleep stages with in decreasing order: 90.5 \% for REM, 85.2 \% for N2, 84.1 \% for Wake and only 67.4 \% for N3 and 63.0 \% on N1. Importantly, this agreement also varies depending on patient, sleep disorders and across sleep centers \cite{Danker-Hopfe2009InterraterStandard} \cite{Rosenberg2014TheEvents}.

Algorithmic approaches have been developed to automatize the process. They are composed of two steps: feature extraction from raw signals and then classification into sleep stages. Among the automated sleep staging methods, we distinguish two main categories: the expert approaches and the deep learning approaches. An expert approach relies on hand-crafted feature extraction followed by a learnt classifier. On the other hand, a deep learning approach learns both the features and the classifier from example epochs.

Numerous studies have focused on expert approaches to classify sleep stages. Spectral and temporal features are computed on raw EEG signals \cite{kerkeni:inria-00112586}  \cite{Berthomier2007} or on multimodal PSG signals \cite{Lajnef2015}. A classifier, like a random forest or a multi-layer perceptron, is then trained on top of these features to estimate the current sleep stage. Most recent approaches take into account successive sleep epochs and feed their features to a recurrent neural network (RNN) to model the time dynamics of sleep \cite{Dong2018}. 

Following the general trend in machine learning, deep learning has also brought new feature extraction methods for automated sleep staging. In \cite{Tsinalis2016a} a convolutional neural network (CNN) extracts relevant features from a single channel raw EEG signal. \cite{Supratak2017} strongly improves the previous approaches by dividing the CNN into two branches to extract features at different scales. A RNN is added after the CNN to model the dependency between contiguous sleep epochs. \cite{Chambon2018} proposed a lighter CNN which can deal efficiently with multimodal data while having fewer parameters than previous methods. \cite{Sors2018} \cite{Cui2018} \cite{Vilamala2017} \cite{Biswal2018} \cite{Fernandez-Varela}  \cite{Stephansen2018} have all reported state-of-the-art performances on various sleep staging datasets with CNN.
These models (excluding \cite{Chambon2018}) have millions of parameters which increases computational cost and the risk of overfitting while lowering data efficiency. Most of these models are applied on a single signal from the PSG which may limit the accuracy of the estimated sleep stages.

\cite{Phan2019} \cite{Phan2018c} \cite{Phan2018b} introduce a different approach, the raw PSG signals of a sleep epoch are transformed into a short term Fourier transform and processed either by a 1D CNN or by a RNN followed by an attention layer \cite{Luong2015}. To model temporal dependencies \cite{Phan2019} feeds the succession of encoded sleep epochs into a second RNN. State-of-the-art performance are reached on the publicly available MASS dataset \cite{OReilly2014}.

Most automated approaches are trained and evaluated on a single manual sleep scoring making it difficult to evaluate how they actually perform considering the low inter-rater agreement. One notable exception, \cite{Stephansen2018} deals with the issue of inter-rater variability using annotations from 6 sleep technologists on a subset of training records. However the multiple sleep staging annotations are not currently publicly available. Another challenge in the evaluation and comparison of automated approaches is that no shared dataset has made a consensus for benchmarking different approaches when it has been shown that performance can greatly vary across datasets \cite{Chambon2018transfer}.
In this study we introduce two publicly available datasets; DOD-H (Dreem Open Dataset - Healthy) and DOD-O (Dreem Open Dataset - Obstructive). DOD-H is built from recordings from 25 healthy adult volunteers. DOD-O is built from recordings from 55 patients suffering from obstructive sleep apnea (OSA). Both datasets were scored by 5 experienced sleep technologists across 3 different sleep centers.
Using these datasets we propose a methodology inspired from \cite{Stephansen2018} and \cite{Arnal662734} to evaluate a sleep stage algorithm against multiple sleep technologists, in order to simulate a real-life setting. This evaluation framework is available at http://github.com/Dreem-Organization/dreem-learning-evaluation together with the scores from the various sleep technologists and the PSG data for both DOD-O and DOD-H.
Using this framework we benchmark and compare several approaches from the literature \cite{Supratak2017} \cite{Phan2019} \cite{Chambon2018} \cite{Tsinalis2016} \cite{Tsinalis2016a}. We also introduce and benchmark a new deep learning method, SimpleSleepNet, inspired by SeqSleepNet \cite{Phan2019}, DeepSleepNet \cite{Supratak2017} and \cite{Chambon2018}. 
First, we compare the performance of human scorers and recent literature models (including SimpleSleepNet) on DOD-H and DOD-O. Then, SimpleSleepNet is used to study the impact on sleep staging performance of the following factors: temporal context, dataset size, number of input signals, size and complexity of the model.
The benchmark code is publicly available at https://github.com/Dreem-Organization/dreem-learning-open.

%% file: 2_materials_and_methods.tex
\section{Materials and Methods}

\subsection{Datasets}

\subsubsection{Dataset 1: healthy patients}
Dataset 1 was collected at the French Armed Forces Biomedical Research Institute's (IRBA) Fatigue and Vigilance Unit (Bretigny-Sur-Orge, France) from 25 volunteers. Volunteers were recruited without regard to gender or ethnicity from the local community via flyers. Volunteers were healthy sleepers without sleep complaints between the ages of 18 and 65, their PSG results confirmed the absence of a sleep disorder. More details and exclusion criteria can be found in \cite{Arnal662734}. Demographics are summarized in Table \ref{tab:demographics}. All participants received financial compensation commensurate with the burden of study participation. The study was approved by the Committees of Protection of Persons (CPP), declared to the French National Agency for Medicines and Health Products Safety, and carried out in compliance with the French Data Protection Act and International Conference on Harmonization (ICH) standards and the principles of the Declaration of Helsinki of 1964 as revised in 2013.
The data used for this study is composed of 12 EEG derivations (C3/M2, F4/M1, F3/F4, F3/M2, F4/O2, F3/O1, FP1/F3, FP1/M2, FP1/O1, FP2/F4, FP2/M1, FP2/O2), 1 EMG signal, left and right EOG signals and 1 electrocardiogram (ECG) sampled at 250 Hz recorded with a Siesta PSG devices (Compumedics).
Each record was scored independently by 5 experienced sleep technologists from 3 different sleep centers following the current AASM Scoring Manual and Recommendations (Version 2.5, Released April 2, 2018). This is based off original 2007 AASM Scoring Manual \cite{Iber2007TheSpecifications}. All scorers are registered Polysomnography Technologists, with over 5 years of clinical / scoring experience.

\subsubsection{Dataset 2: patients with OSA}
The dataset 2 was collected at the Stanford Sleep Medicine Center and consists of PSG recordings from 55 patients (Clinical trial number NCT03657329). Patients were included in the study based on clinical suspicion for sleep-related breathing disorder. Individuals with a diagnosed sleep disorder different from OSA were excluded from this study. Exclusion criteria can be found in \cite{thorey2019}. Demographics are given in Table \ref{tab:demographics}. All trial participants gave their informed written consent prior to participation. They received compensation for their participation.
The data used for this study is composed of 8 EEG derivations (C3/M2, C4/M1, F3/F4, F3/M2, F4/O2, F3/O1, O1/M2, O2/M1), 1 EMG derivation, left and right EOG signals and 1 electrocardiogram (ECG) sampled at 250 Hz recorded with a Somno HD PSG devices (Somnomedics).
Each record was scored independently by 5 experienced sleep technologists from 3 different sleep centers following the current AASM Scoring Manual and Recommendations (Version 2.5, Released April 2, 2018). This is based off original 2007 AASM Scoring Manual \cite{Iber2007TheSpecifications}. All scorers are registered Polysomnography Technologists, with over 5 years of clinical / scoring experience.

\input{tables/demographics.tex}

\subsection{Evaluation in the context of multi-scoring}
\label{sec:multiscoring}

The process of evaluating the performance of a human scorer, or an automated approach, against a consensus of multiple human scorers is inspired from \cite{Stephansen2018} and has been presented in our previous work \cite{Arnal662734}. The goal is to use reduce the known inter scorer variability for sleep stage classification by using a majority vote from the sleep experts. In this section, we highlight the main aspects and differences.

\newcommand{\varB}[1]{{\operatorname{\mathit{#1}}}}
\subsubsection*{Soft-Agreement}\hfill\\
When taking a majority vote between sleep experts, ties can occur. In case of ties, we choose to use the label of the most reliable scorer. The most reliable scorer will be the one that is the most in agreement with all the other scorers on a record. To find this scorer, we defined Soft-Agreement in \cite{Arnal662734} as follows.
\textit{Notations : }Let $y_j \in \llbracket 4 \rrbracket^{T}$ be the sleep staging associated to scorer $j$ taking values in $\{ 0, 1, 2, 3, 4 \}$ standing respectively for Wake, N1, N2, N3 and REM with size $T$ epochs. Let $N$ be the number of scorers. Let $\hat{y}_j \in \{ 0, 1 \}^{5 \times T}$ be the one hot encoding of $y_j$.

To evaluate a sleep staging of one record against multiple sleep staging methods, we introduced in \cite{Arnal662734} a Soft-Agreement metric defined as:

\begin{equation*}
        \varB{Soft-Agreement}_j = \frac{1}{T}\sum_{t=0}^{T}\hat{z}_j[y_j]
\end{equation*}

with:
\begin{equation*}
    \hat{z}_j[t] = \frac{\sum\limits_{\substack{i = 1 \\ i \neq j }}^{N}\hat{y}_i[t]}{max \sum\limits_{\substack{i = 1 \\ i \neq j }}^{N}\hat{y}_i[t]} \forall t
\end{equation*}

This metric measures how close the sleep staging of interest is from all the other scorers sleep staging. It values 1 if the sleep staging of interest is always in agreement with the majority vote (or one of the majority votes in case of ties).

\subsubsection*{Other metrics} \label{explanation_consensus} \hfill\\
To merge multiple sleep stagings into a single consensus sleep staging, we simply take the majority vote on each 30-second epoch. When a tie occurs on a specific epoch, we take the sleep stage scored on the sleep staging with the highest Soft-Agreement on the record. This differs from our previous work \cite{Arnal662734} where we used the scorer with the highest soft-agreement over all the records of the dataset, hence inducing a dependency to the dataset. We also compute a weight between 0 and 1 for each epoch based on how many scorers voted for the consensus sleep staging epoch. These weights are used to balance the importance of each epoch in the computation of each of the following metrics.

To measure agreement between two sleep stagings on a specific record, we measure  $\varB{F1-score} = 2*\frac{Pr * Re}{Pr + Re}$ with $Pr = \frac{TP}{TP + FP}$ and $ Re = \frac{TP}{TP + FN}$, and TP, FP, and FN are the number of true positives, false positives, and false negatives, respectively. The score is computed per-class, one class against the others, and averaged taking the proportion of each class into account. We also provide \textit{Accuracy}, as the ratio of correct answers and \textit{Cohen's Kappa}, $\kappa = \frac{p_j - p_e}{1 - p_e}$ where $p_j$ is the relative observed agreement and $p_e$ is the hypothetical probability of chance agreement.

\subsection{SimpleSleepNet}
\label{sec:simple_sleep_net}

\begin{figure*}
\begin{center}
  \includegraphics[width=0.955 \textwidth]{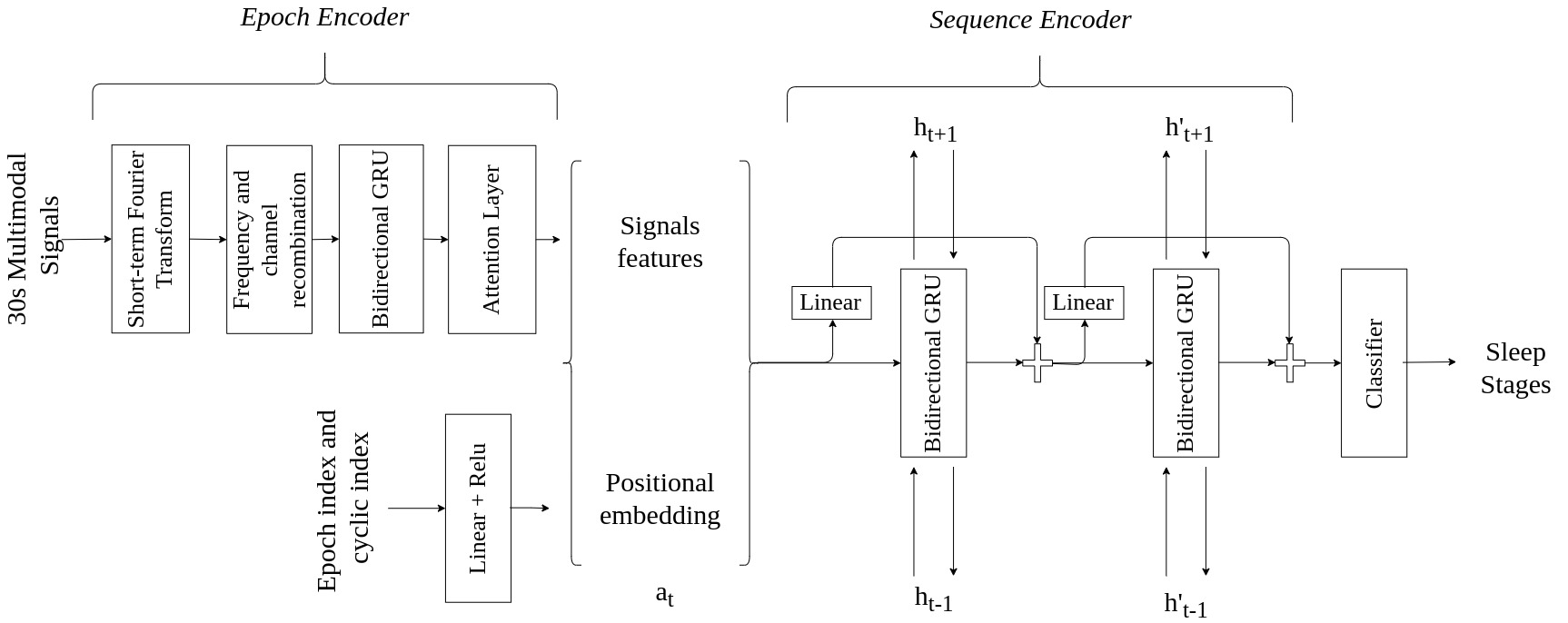}
  \caption{\textbf{SimpleSleepNet overview diagram:} 
  $h_{t-1}, h'_{t-1}$ represent the hidden state from the previous epoch of the sequence and $h_{t+1}, h'_{t+1}$ the hidden state from the next epoch of the sequence. $a_t$ is the embedding of the current epoch.}
  \label{SimpleSleepNetFigure}
\end{center}
\vspace{-1.8em}
\end{figure*}

SimpleSleepNet is a new automated sleep staging model based on recent advances in the field. The initial stage in SimpleSleepNet is inspired by \cite{Phan2019}. Where it differs from the latter is in its use of a channel-wise dropout. Moreover, it replaces the filter bank with a linear layer, recombines the EEG derivations using a linear layer (an approach inspired by  \cite{Chambon2018}), and omits the norm from the GRU. The second stage, which models epoch dependencies, is inspired by \cite{Supratak2017} but differs from it in that it uses positional embedding. SimpleSleepNet uses fewer parameters than the other models by reducing the size of the hidden layers.
In this section, a comprehensive description of each module of SimpleSleepNet is presented. Figure \ref{SimpleSleepNetFigure} summarizes the overall architecture of the network.

\subsubsection{Spectrogram}
The short-term Fourier transform (STFT)  is computed on the preprocessed signals of each of the epochs. Preprocessing is defined in section \ref{experiment:setup}. Each epoch is in $\mathbb{R}^{C, 30 \cdot  fs }$ where $C$ denotes the number of channels and $fs$ the signal frequency. During training, signals are randomly set to zero before computing the STFT with a probability $p_{kill}$ to reduce overfitting.  

Similarly to \cite{Phan2019}, the STFT is computed over 256 points of signal every one second with a Hanning window. The log-power of the STFT is taken and clipped between -20 and 20. Each epoch is thus represented by a time-frequency picture $\mathbf{S} \in \mathbb{R}^{C,T,N}$ where $C$ is the number of signals, $T = 28$ the number of time-steps  and $N = 129$ the number of frequency bins. The clipped STFT is 0-mean 1-variance normalized signal-wise independently of the timestep. Mean and variance are computed over all the training records.

\subsubsection{Signals and frequencies reduction}
First the $N$ frequency bins are linearly reduced into $n \leq N$ filters, and the $C$ input signals are linearly reduced into $c \leq C$ signals. Their weights matrices are respectively in $\mathbb{R}^{n,N}$ and $\mathbb{R}^{c,C}$ and their bias in $\mathbb{R}^n$ and $\mathbb{R}^c$. The linear projections are applied respectively along the frequencies and signals axis to project the initial spectrogram from $\mathbb{R}^{C,T,N}$ into $\mathbb{R}^{c,T,n}$
The two projections are applied independently. Dropout is then applied with a probability $p_1$.
\subsubsection{GRU with attention}
The recombined signals are reshaped into $\mathbb{R}^{T,c.n}$ and fed to a bidirectional Gated Recurrent Unit (GRU) \cite{GRU2014} with $m_1$ hidden units to build a representation in $\mathbb{R}^{T,2.m_1}$. Dropout is applied after the GRU with the same probability $p_1$.
Then, the output of the GRU is fed into an attention layer. The attention layer is implemented as presented in \cite{Luong2015} with context size $m_{ctx}$. The attention layer reweights and sums the GRU hidden states along the time axis to build a vector representation of the current sleep epoch in $\mathbb{R}^{2.m_1}$

\subsubsection{Positional embeddings}
Positional embeddings have recently been used in Transformer architectures \cite{transformer2017} to model time dependency. Here, positional embedding is used to include global context in the sequential modelling layer. The positional embedding of an epoch is composed of the scaled index epoch $i^{epoch}_t$ and of five cyclic indexes $i^{cycle}_{t,l}$ where $t$ is the number of sleep epochs since the beginning of the night. Then $i^{epoch}_t= \frac{t}{1200}$ and $i^{cycle}_{t,l}=cos(\frac{t.\pi}{l}) \text{ for $l$ in [30, 60, 90, 120, 150]}$.
The concatenation $[i^{epoch}_t,i^{cycle}_{t,30}, \ldots,i^{cycle}_{t,150}] \in \mathbb{R}^{6} $ is then projected, using a linear layer with weights and bias in $\mathbb{R}^{6,6}$ and $\mathbb{R}^6$, to build $i_t$.
Then, $i_t$ is concatenated with the output of the attention layer to compute the current epoch representation $a_t \in \mathbb{R}^{2.m_1 + 6}$

\subsubsection{Sequence encoder and classifier}
Given a temporal context $k$ and a central epoch $t$, the epochs $a_{t-k}, ..., a_{t+k}$ are fed to a two layers bidirectional GRU with skip-connections (SkipGRU) and $m_2$ hidden units.
The SkipGRU is similar to the sequence encoder of DeepSleepNet \cite{Supratak2017} with additional intermediary skip connections. Given its input size $2.m_1 + 6$, the SkipGRU has a weights matrix $\mathbf{W}_{skip} \in \mathbb{R}^{m_2,2.m_1 + 6}$ and a bias vector $b_{skip} \in \mathbb{R}^{m_2}$ and follows:
$$ h_t = \frac{1}{2} [GRU(a_t,h_{t-1}) + \mathbf{W}_{skip} a_t + b_{skip}]$$ 
The bidirectional SkipGRU is built by concatenating the outputs of a forward and of a backward SkipGRU. Dropout is applied on $h_t$ with a probability $p_2$. We denote $h_{t-k}, \ldots, h_{t+k} \in \mathbb{R}^{2\cdot k_2}$ its outputs.\\
This sequence is fed to a final softmax classification layer which outputs the sleep stages probabilities $\hat{\pi}^{(t)}_{-k}, \ldots, \hat{\pi}^{(t)}_{k} \in \mathbb{R}^{5}$ 

\subsubsection{Loss function}
Since SimpleSleepNet outputs several sleep stages estimates instead of a single one, the loss has to be modified accordingly (similarly to \cite{Phan2019}). Let $\mathbf{S}=[s_{t-k},\ldots, s_{t+k}]$ be the input sequence of the spectrograms from $2k+1$ sleep epochs. For the epoch t, the loss is defined as $\mathcal{L}(\mathbf{S},y) = -\frac{1}{2k + 1} \sum_{i=-k}^{k} \hat{y}_{t+i}  \boldsymbol{\cdot} log(\hat{\pi}^{(t)}_{t+i}(\mathbf{S}))$

\subsection{Evaluation}  \label{section_evaluation}
At evaluation time, the multiple available predictions for an epoch are aggregated following \cite{Phan2019}: given an epoch $t$ and a temporal context $k$, the aggregated sleep stage probabilities is the geometric mean 
$\tilde{\pi}^{(t)} = \text{exp}\left( \frac{1}{2k + 1} \sum_{i = -k}^{i =k} log(\hat{\pi}^{(t+i)}_{t}) \right)$
and the predicted sleep stage used for evaluation is $\tilde{y}^{(t)} = \text{argmax}_{\; j \in [\![0,5]\!]} \tilde{\pi}^{(t)}_j $

%% file: tables/demographics.tex
\begin{table}[H]
\centering
\scalebox{1.2}{%
\begin{tabular}{|ccc|}
 \hline
                          & DOD-H       & DOD-O       \\
 \hline
Age                       & 35.32 $\pm$ 7.51 & 45.6 $\pm$ 16.5 \\
M/F                       & 19 / 6        & 35 / 20        \\
BMI                       & 23.81 $\pm$ 3.43  & 29.6 $\pm$ 6.4 \\
AHI                       & -           & 18.5 $\pm$ 16.2  \\
Sleep Time (min)    & 427         & 387          \\
Sleep Efficiency (\%)     & 86.4        & 79.7         \\
Sleep Onset Latency (min) & 19.5        & 19.7         \\
Wake (\%)                 & 13.4        & 20.0           \\
N1 (\%)                   & 7.10        & 6.11          \\
N2 (\%)                   & 46.7        & 46.8         \\
N3 (\%)                   & 14.4        & 11.9         \\
REM (\%)                  & 18.2        & 14.8       \\
 \hline
\end{tabular}%
}
\caption{Demographics for DOD-H and DOD-O. More information can be found here \cite{Arnal662734} for DOD-H and \cite{thorey2019} for DOD-O. All values are average across all subjects.}
\label{tab:demographics}
\vspace{-2em}
\end{table}

%% file: 3_experiments.tex
\section{Experiments}

\subsection{Baselines}
\label{sec:baseline}

To benchmark the current state-of-the-art in automated sleep staging on both DOD-O and DOD-H, we selected recent approaches from the literature reporting good performances on publicly available datasets. These approaches were reimplemented in Pytorch \cite{paszke2017automatic}, for reproducibility the code is publicly available in the following repository: https://github.com/Dreem-Organization/dreem-learning-open. The presented approach SimpleSleepNet is also included in the benchmark.

\subsubsection{Mixed neural network (Expert approach) \cite{Dong2018}}

The Mixed neural network (MNN) computes aggregated features (average, median, maximum, minimum, standard deviation, entropy) on the raw signal. The aggregation is performed on the complete epoch and on sliding windows of 5 seconds with 3.5 seconds of overlap. Similarly, time-frequency features are computed using the Fourier transform over windows of 5 seconds with 3.5 seconds of overlap and on the complete epoch. The amplitude of the Fourier transform is summed over frequency bands of interest for sleep, general statistics are computed for each epoch and for each band and are used as additional features. The computed features are fed to a two-layer, fully-connected neural network (FCNN) with dropout and then to a bidirectional LSTM followed by a classification layer. The features are computed on the F4-M1 derivation on DOD-H dataset and on F4-O2 on DOD-O.

\subsubsection{Tsinalis et al. \cite{Tsinalis2016a}}
Tsinalis et al. \cite{Tsinalis2016a} introduced the first CNN for sleep staging. The model takes 630 seconds of raw signals (which is equivalent to 21 sleep epochs) centered on the current epoch. The signal is fed to two successive convolution + pooling layers with Relu activations. The features are then flattened and fed to a two-layer FCNN followed by the classification layer. The network estimates the sleep stage of the central epoch. The parameters are those provided in the original paper. However, for a fair comparison with the other models, the net is trained on all the PSG signals instead of the single channel without any other architectural change.

\subsubsection{Chambon et al. \cite{Chambon2018}}
Chambon et al. \cite{Chambon2018} built a convolutional model to handle multivariate and multi-modal signals. The model uses 630 seconds (21 sleep epochs) of signals as its input, the model classifies the central epoch. First a convolution of size 1 is applied, the convolution does not take into account the time and is only applied over the signals. This convolution models the dependencies between the different signals to learn virtual signals which are good representations of the original signals. Then a succession of two Convolution and Pooling layer blocks is applied on each virtual signal independently. Processing each signal independently reduces the overall complexity and increases the inference and training speed. The output of the CNN is flattened before being fed to a final classification layer. The parameters are the one used in the original paper, the net is trained on all the PSG signals. 

\subsubsection{DeepSleepNet \cite{Supratak2017}}
DeepSleepNet improves  \cite{Tsinalis2016a} with a hierarchical model, first, each epoch is encoded, then the succession of the epochs is processed by a recurrent network to model temporal dependencies.
Instead of having only one convolutional layer, each sleep  epoch is encoded by two distinct convolutional networks with different filters and pooling sizes. The first network has smaller filter sizes and is focused on temporal information while the second network has a larger filter size and focuses on frequency information. The output of both networks are concatenated to build the representation of the epoch. To deal with the stage transition, a succession of 2 bidirectional LSTM with a skip-connection processes the sequence of encoded sleep epochs. The model is trained on all the signals.

\subsubsection{SeqSleepNet \cite{Phan2019}}
 SeqSleepNet takes the spectrogram of the signal as the input, the number of Fourier bins is reduced with a learned frequency filter-bank which projects the original bins on a smaller frequency space. The reduced STFT is then fed to a bidirectional LSTM with recurrent batch-normalization \cite{cooijmans2016recurrent} followed by an attention layer. The attention layer reduces the temporal dimension and encodes the 30-second sleep epoch into a single vector. The encoded representations of consecutive sleep epochs are then fed to a bidirectional GRU, the output of the GRU is used by the classification layer to output the final sleep stage estimate.

\subsubsection{SimpleSleepNet}
The Fourier bins are projected on $n=30$ filters and the original number of channels is kept ($c=C$).The dropouts probabilities $p_{kill}, p_1, p_2$ are set to 0.5. $m_1=m_2=25$ hidden units are used in both the epoch encoder and the sequence encoder. The attention context size $m_{ctx}$ is also set to 25.

\subsection{Benchmark setup}  \label{experiment:setup}
\label{benchmark-setup}
$\varB{Soft-Agreement}$ was computed for all scorers on all records. Following \ref{explanation_consensus} we used these values to build a consensus hypnogram for every record. The human scorers are individually evaluated against the consensus hypnograms built from the four others. The automated approaches are trained and evaluated with the consensus hypnograms built from the four overall best scorers in terms of overall best $\varB{Soft-Agreement}$. On DOD-H the 5 human scorers had an overall $\varB{Soft-Agreement}$ of respectively 0.87, 0.91, 0.92, 0.84 and 0.92 so scorers 1, 2, 3 and 5 are selected. On DOD-O, the 5 human scorers had an overall $\varB{Soft-Agreement}$ of 0.88, 0.87, 0.88, 0.88 and 0.91 respectively, so scorers 1, 2, 4 and 5 are selected. In practice, ties occurred on average for 7.3 \% of the epochs in DOD-H and 9.9 \% of the epochs in DOD-O.\\
The same preprocessing is used for all the models, a band-pass filter is applied between $[0.4,18] Hz$ to remove residual PSG noise, then, the signals are linearly resampled at $fs = 100 Hz$ to reduce the training computational cost. Each signal is then clipped and divided by 500 to remove extreme values. Predictions on each epoch are computed using a temporal context of past and future epochs (see section \ref{section_evaluation}). To ensure having points for the very first and last epochs of the record, a zero-padding with at least the same length as the temporal context is added at the start and end of each record.  \\
The models are trained using back propagation with the Adam optimizer and a learning rate of 0.001, momentum parameters $\beta_1=0.9$ and $\beta_2=0.999$ and a batch size of 32. All the models are trained for a maximum of 100 epochs with early stopping. The training was stopped when validation accuracy stopped improving for more than 15 epochs. The model with the best validation accuracy is used to evaluate the model. The temporal context is set to 21 for all the models, increased temporal context has been shown to improve performances in \cite{Chambon2018} and \cite{Phan2018c}. Hence, using the same temporal context ensures the benchmark fairness. Furthermore for each model from the literature, several set of hyper-parameters were evaluated on DOD-O and DOD-H, the best run is reported for these models. \\
On DOD-H the models were evaluated in a leave-one-out way: 18 records are used for training, 6 are kept for validation and 1 is kept to test the model. On DOD-O the models were evaluated in a 10-folds validation way: 37 records are used for training, 12 are used for validation and 6 records to evaluate the model.\\
The number of parameters of each model and training time for one epoch on a Titan-X on DOD-O are given for reference Table \ref{tab:parameters_count}.

\input{tables/parameters_count.tex}

\subsection{Benchmark on DODO and DODH}

\begin{figure*}[h]
    \centering
    \begin{minipage}{.5\linewidth}
    \begin{center}
    \footnotesize{DOD-H dataset}
    \includegraphics[width=.98\linewidth]{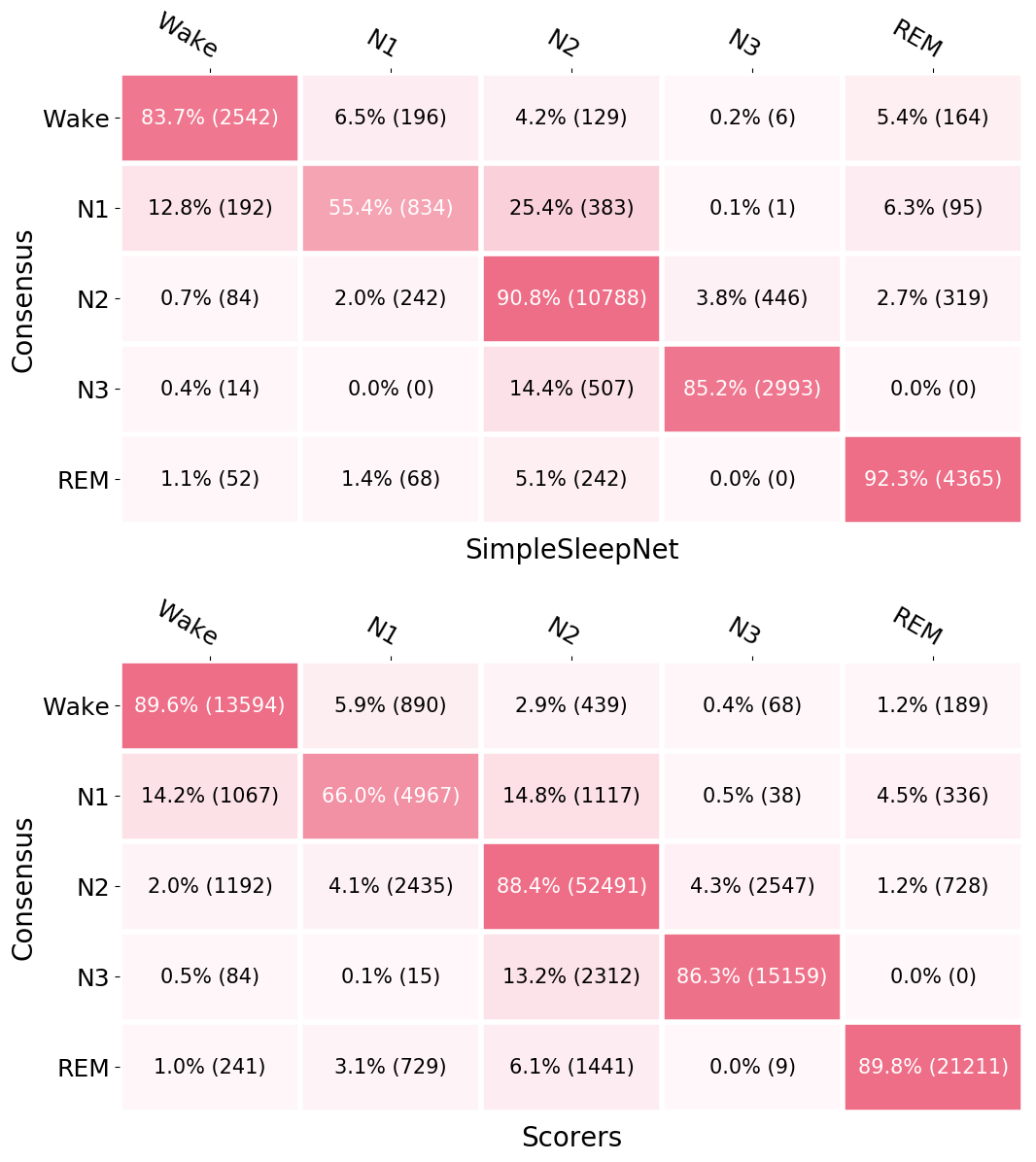} 
    \end{center}
    \end{minipage}%
    \begin{minipage}{.5\linewidth}
    \begin{center}
    \footnotesize{DOD-O dataset}
    \includegraphics[width=.98\linewidth]{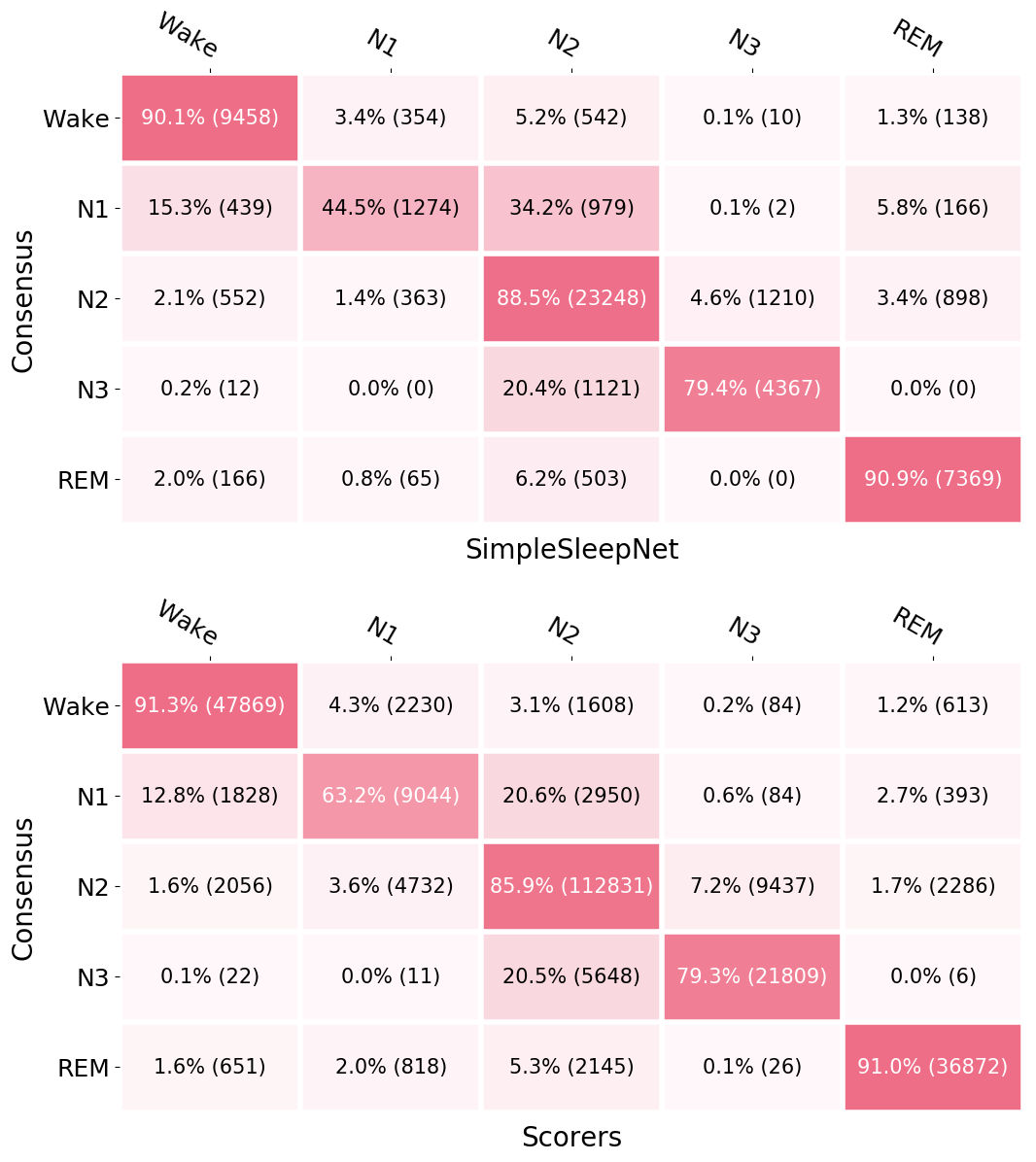} 
    \end{center}
    \end{minipage}
    \caption{Confusion matrix for SimpleSleepNet versus consensus hypnograms built from the top four best scorers (top) and the overall confusion matrix for human scorers versus the consensus hypnograms built from the four other scorers  (bottom) for DOD-H (left) and DOD-O (right).  Values are normalized by row with and the number of epochs is given in parentheses.}
    \label{fig:confusion}
\vspace{-1em}
\end{figure*}

The overall, best and worst performances of the five scorers are reported in Table \ref{tab:complete_results} for both datasets, all the metrics are computed subject-wise and not epoch-wise to be more representative of a clinical setting. On DOD-H, the average scorer F1 is 86.8 $\pm$ 7.6 \%. The average scorer accuracy is above the one reported in \cite{Danker-Hopfe2009InterraterStandard}. F1 is higher for REM (90.8 $\pm$ 10.3 \%), followed by N2 (88.9 $\pm$ 7.6 \%), Wake (84.3 $\pm$ 13.6 \%) and lower for N3 (78.5 $\pm$ 23.9 \%) which also shows the highest variability. N1 has the lowest F1 (50.3 $\pm$ 14.7 \%).\\
On DOD-O, the performances of the scorers are slightly lower than on DOD-H with an overall scorer F1 of 84.8 $\pm$ 8.6 \%. F1 is higher for Wake epochs (90.8 $\pm$ 8.2 \%). For all the other stages it is slightly lower with 85.6 $\pm$ 23.3 \% for REM, 85.6 $\pm$ 10.7 \% for N2 and 44.6 $\pm$ 16.8 \% for N1. N3 is notably lower with an F1 of 56.9 $\pm$ 33.1 \%. Standard deviation (SD) sensibly increases for all the stages compared to DOD-H. Figure \ref{fig:confusion} shows the scorers confusion matrices on both dataset, most of the errors involve N1 being mistaken for WAKE or N2 and N3 being mistaken for N2.\\
The performances of the automated approaches are also given in Table \ref{tab:complete_results}. SimpleSleepNet shows the best performance on both datasets for the considered metrics when compared to both humans and other approaches. On DOD-H, SimpleSleepNet is better than the best scorer and shows a lower SD with an F1 of 89.9 $\pm$ 4.1 \%. On DOD-O, it also performs better but with a slightly higher SD than the best scorer with an F1 of 88.3 $\pm$ 9.0 \%. With the exception of \cite{Chambon2018} and \cite{Tsinalis2016a}, every model performs better  with a much lower variability on DOD-H than on DOD-O. Except \cite{Tsinalis2016a}, most models have F1 scores which are on par with the scorers' average and above the worst scorer.

\input{tables/complete_results.tex}

\subsection{SimpleSleepNet ablation study}
To assess the importance of each of the modules of the architecture of SimpleSleepNet, ablated models were trained on both datasets.
While technically not being an ablation of the model itself, the influence of the preprocessing step is assessed in \textit{No filtering} where the filtering is removed. In \textit{No channel dropout} the channel dropout is removed ($p_{kill} = 0$). Then, we evaluate the effects of the blocks of the epoch encoder. In \textit{No frequency reductions} the linear frequency reduction is removed, in \textit{Filter bank} it is replaced by a filter-bank \cite{Phan2019}, in \textit{No channel recombination} the linear channel recombination is removed and in \textit{No attention} the attention layer is replaced by an average-pooling layer. The architecture of the sequence encoder is analyzed by removing the positional embedding in \textit{No positional embedding}, by using a single layer in the GRU encoder in \textit{Single GRU layer}, and by removing the skip-connection in \textit{No skip connection}. 

\input{tables/ablations.tex}

The results are shown in Table \ref{tab:ablations}. 
Removing the frequencies reduction layer or the channel dropout are the most impacting ablations on both datasets. Other ablations do not significantly impact the performance on DOD-H. However, on DOD-O, the filtering and the filter bank greatly impact the performance. Other ablations also demonstrate the slight improvement provided by each layer on DOD-O. Overall, the full model presents the best ranking on both datasets.

\subsection{Influence of the experimental setup}

\subsubsection{Model size}
To assess the influence of the model size on performances, two variants of SimpleSleepNet are evaluated SimpleSleepNet-Small and SimpleSleepNet-Large. SimpleSleepNet-Small  (resp. SimpleSleepNet-Large) has hidden units of size $m_1 = m_2 =12$ (resp. $m_1= m_2 =50$) in both GRU and the attention layer context size is set to $m_{ctx} = 12$ (resp. $m_{ctx} =50$). SimpleSleepNet-Small has approximately three times less parameters and SimpleSleepNet-Large three times more parameters than SimpleSleepNet as show in Table \ref{tab:parameters_count}. 

\input{tables/model_size.tex}

Increasing the model size increases SimpleSleepNet performances both on DOD-O and DOD-H as shown in Table \ref{tab:model_size}. On DOD-H F1 increases by 0.5 \% for the large model and is reduced by 0.6 \% for the small model. On DOD-O, F1 is increased by 0.7 \% with the large model and reduced by 1.1 \% when using the small model. On both datasets, using larger models reduces variance significantly. 

\subsubsection{Performances on a single EEG derivation}
We assess the performance of SimpleSleepNet on a the F4-O2 derivation on both datasets in Table \ref{tab:single_channel}. Performances are significantly lower compared with a model trained on the full montage, the single channel model F1 score is 3.9 \% points lower on DOD-O and  3.3 \% points lower on DOD-H. The model F1-score with single channel is still on par with the scorers average.

\input{tables/single_channel.tex}

\subsubsection{Size of the training set}
\begin{figure}
\vspace{-2em}
    \begin{minipage}{.48\linewidth}
    \includegraphics[width=0.98\linewidth]{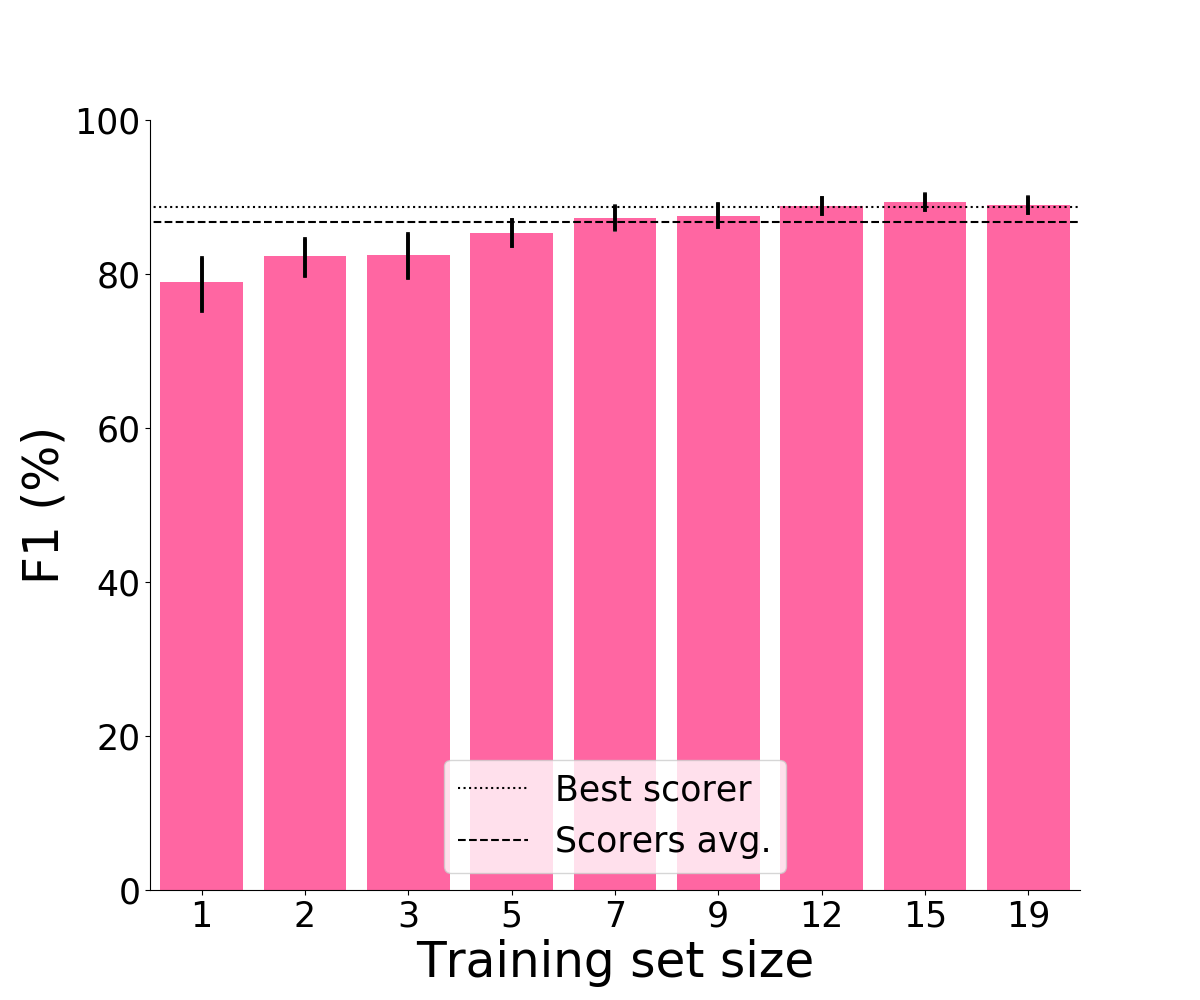} 
    \end{minipage}
    \begin{minipage}{0.48\linewidth}
    \includegraphics[width=0.98\linewidth]{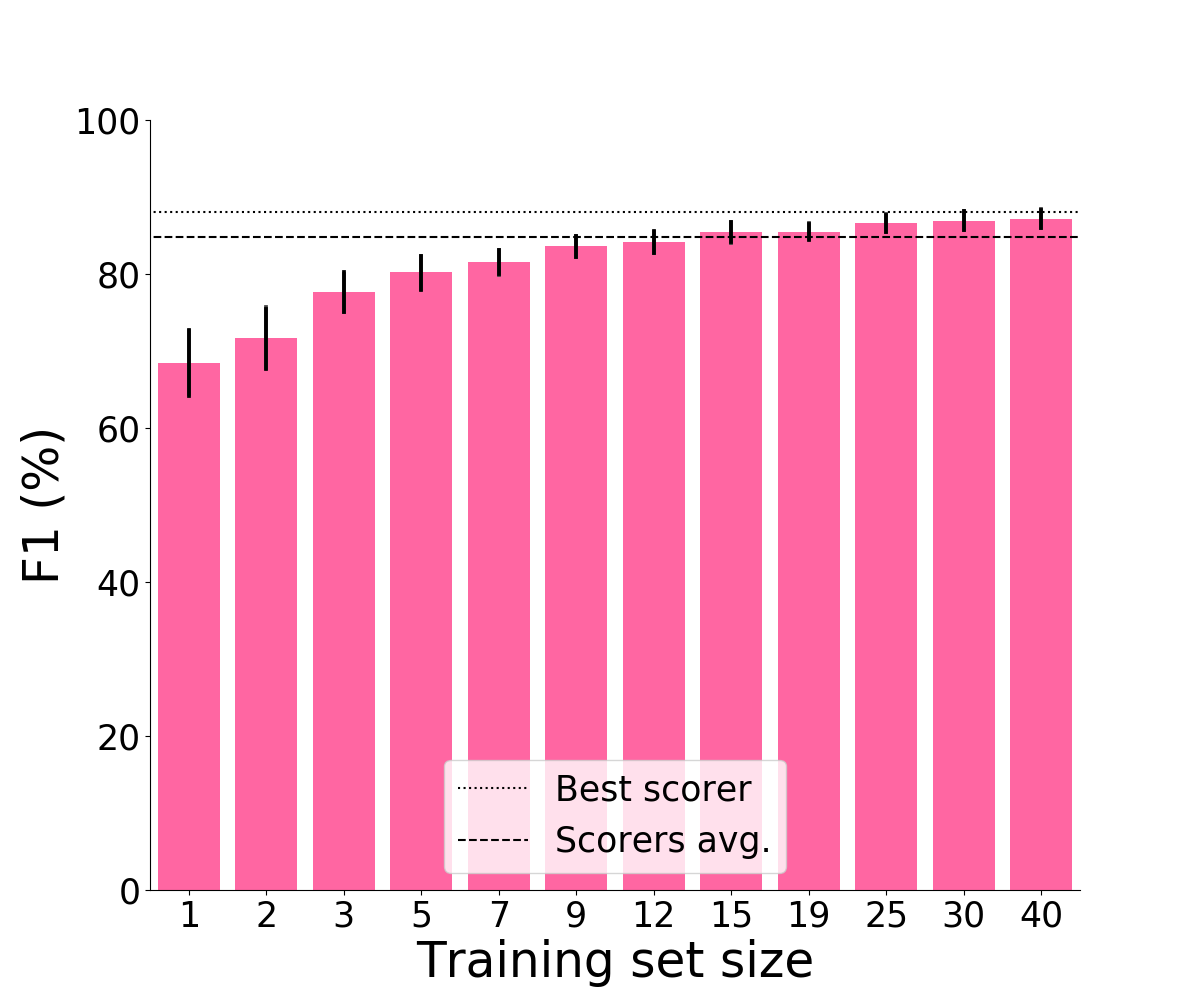} 
    \label{learning_curve_stanford}
    \end{minipage}
    \caption{Evolution of the F1 w.r.t the training set size on DOD-O (right) and DOD-H (left) dataset.}
    \label{learning_curve_irba}

\end{figure}

Labelling records is a costly and long process, hence having data efficient models is crucial. To assess the data efficiency of SimpleSleepNet, the model was trained with training set of increasing size $k$ (1 to 19 for the DOD-H dataset, and 1 to 40 for the DOD-O dataset). For a given training repetition, the split is built in the following way for DOD-H (resp. DOD-O), first 3 (resp. 5) records are randomly sampled for the validation set and 3 (resp. 5) records are sampled for the test set. Out of the 19 (resp. 45) remaining records, the training set of size $k$ is built with the first $k$ records. This experiment is repeated 20 times. The mean F1 and the 95 \% confidence interval on the test set are computed over the 20 experiments are presented Figure \ref{learning_curve_irba}.
Human level performances are reached on both datasets with less than 20 records, DOD-O has a steeper learning than DOD-H. On DOD-H  the F1 reaches a plateau where incremental gains are low with 12 records, while 25 records are required to reach a plateau on DOD-O . The average scorer performance is reached with 7 records (resp. 15). In addition to the increased F1, the standard deviation of the test F1 strongly decreases with the number of training records. 

\subsubsection{Temporal context}

\begin{figure}
    \begin{center}
    \begin{minipage}{.5\linewidth}
    \begin{center}
    \includegraphics[width=\linewidth]{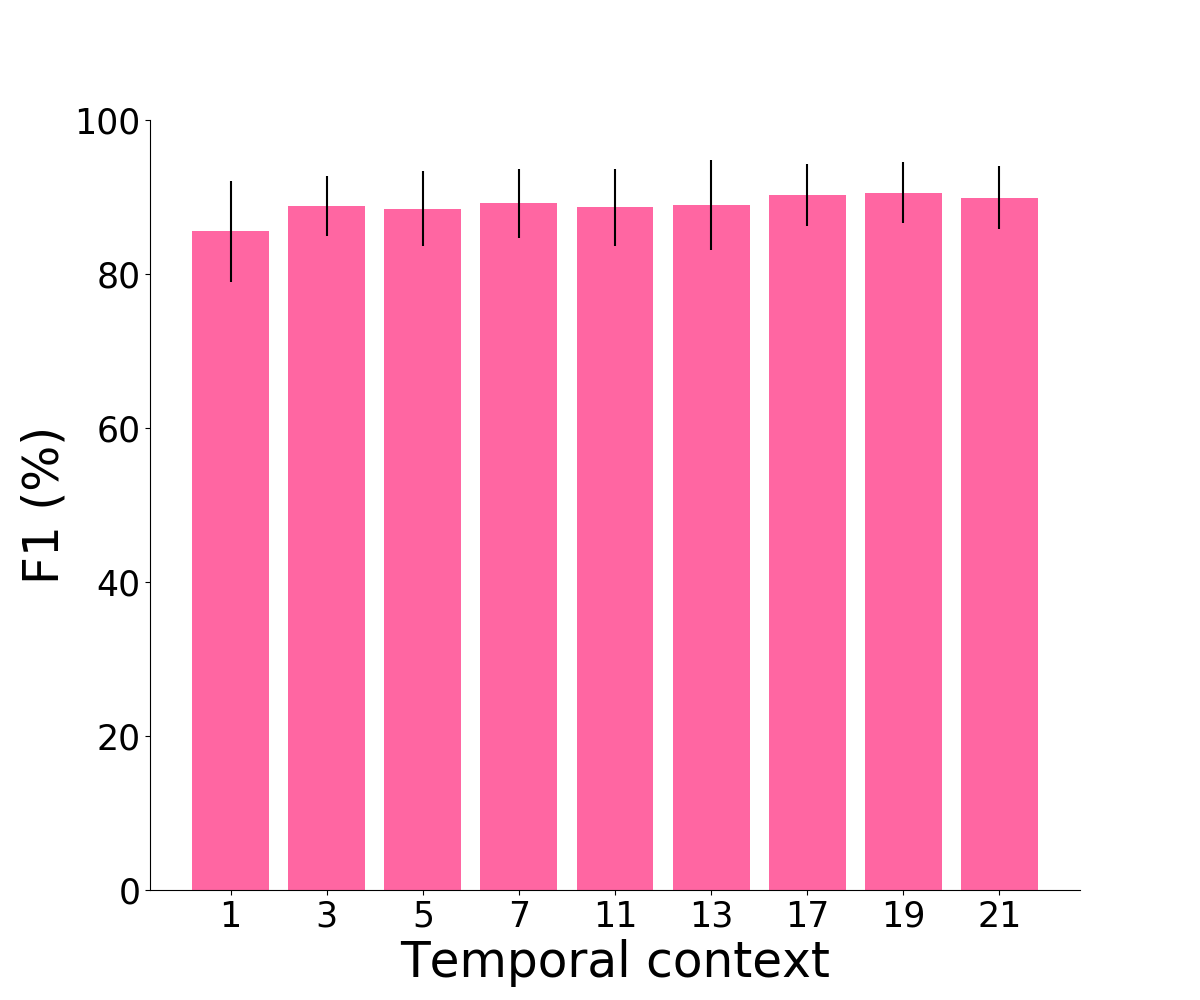} 
    \end{center}
    \end{minipage}%
    \begin{minipage}{.5\linewidth}
    \begin{center}
    \includegraphics[width=\linewidth]{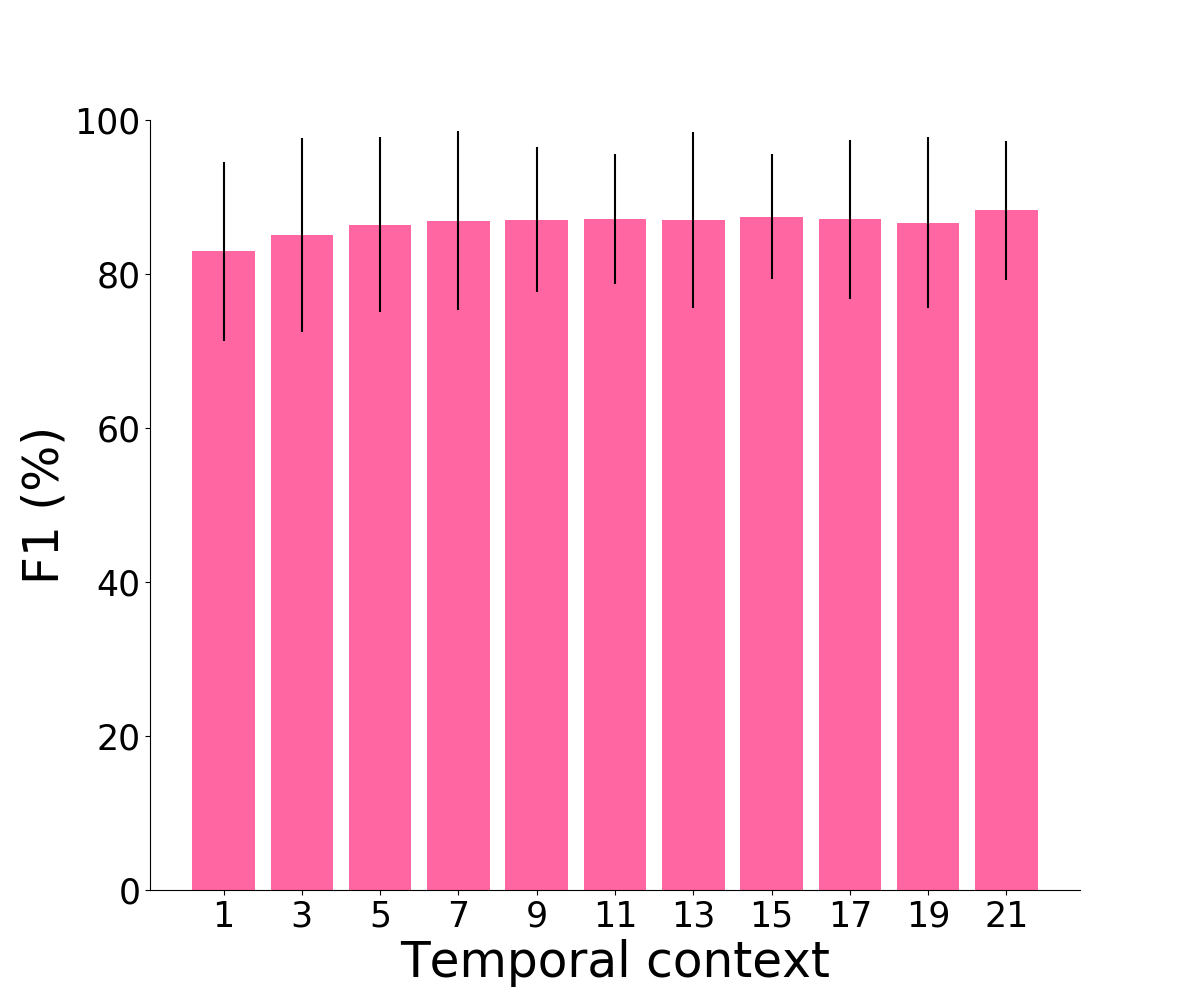} 
    \end{center}
    \end{minipage}
    \caption{Evolution of the F1 w.r.t the temporal context on DOD-H (right) and DOD-O (left) dataset.}
    \label{temporal_context}
    \end{center}
\vspace{-1em}
\end{figure}

We study the impact of the temporal context on the performance by training SimpleSleepNet on DOD-H and DOD-O with varying sizes of temporal context. The size of the temporal context is incrementally increased from 1 (no temporal context apart from the current epoch) to 21 (ten epochs before and after the current epoch). Results are presented Figure \ref{temporal_context}. Even with a single epoch, performances are decent on both dataset with a F1 of 85.5 $\pm$ 6.5 \% on DOD-H and 83.0 $\pm$ 11.6 \% on DOD-O. The F1 sensibly increases when the temporal context is increased from 1 to 7, then it plateaus.

\subsection{Direct transfer learning}
In a real-life, clinical setting, one may wish to train a staging model of a source dataset and to use it on another unlabelled dataset. To assess the transferability of SimpleSleepNet, we train and validate it on DOD-H (resp. DOD-O) and test it on DOD-O (resp. DOD-H). The experiment is repeated 20 times, for each repetition, 70 \% of the records from the source dataset are randomly selected for training and the remaining 30 \% for validation. All the records of the target dataset are used to test the model performance.

\input{tables/transfer_learning.tex}

The results of the experiment are shown in Table \ref{tab:transfer_learning}. When SimpleSleepNet is trained on DOD-O and evaluated on DOD-H, the F1 drops from 89.9 \% to 84.8 \% compared to a model trained from scratch on DOD-H. The standard deviation of the performance metrics almost doubles. The performance drop is bigger when the model is trained on DOD-H and evaluated on DOD-O, the F1 drops from 88.3 \% to 62.6 \%.

\subsection{Benchmark on external dataset}

\subsubsection{MASS SS3 \cite{OReilly2014}}
The MASS SS3 cohort is composed of 62 nights from healthy subjects, done with a full PSG montage (20 scalp EEG,2 EOG, 3 EMG and 1 ECG) and manually scored by a sleep expert according to the AASM standard. The models were trained on the C4-O1, F4-EOG Left, F8-Cz, on the average of the two EOGs and on the average of EMG-Chin1 and EMG-Chin2 which are available for all records and frequently used by the models evaluated on MASS. We used the same preprocessing and training parameters as in the previous section \ref{benchmark-setup}. The models are evaluated in a 31-folds validation way (as in \cite{Supratak2017}).

\subsubsection{Sleep EDF \cite{sleep_edf}}
The Sleep EDF database contains 197 nights from 106 subjects, amongst these nights, 153 are from 82 subjects without any sleep-related medications (SC study) and 44 are from subjects with trouble falling asleep (ST study). 22 of the 44 nights are done after a Temazepam intake. We consider two splits, S-EDF-20 with the subjects 0 to 19 from the SC study and S-EDF-Extended will all the subjects from the database. Similarly to \cite{Supratak2017} \cite{utime}, we only considered the epochs in-between 30 minutes before the first non-wake epoch epoch and 30 minutes after the last non-wake epoch. The models are trained and evaluated using a 20-folds CV on S-EDF-20 and 10-folds CV on S-EDF-Extended. Records from a subject are in the same fold.
The models are trained on the FPZ-Cz, Pz-Oz and the EOG derivation without further processing.

\input{tables/external_datasets.tex}

\subsubsection{Results} The results are presented Table \ref{tab:mass_evaluation}. Our implementation of the literature models reaches equal or improved performance when compared to the original publications. This improvement can be explained by three different reasons. First, we used more derivations than in the original papers. \cite{Tsinalis2016a},\cite{Tsinalis2016},\cite{Supratak2017} used a single derivation and \cite{Phan2019} three-derivations. Secondly, the prediction from a single epoch is the average of the prediction over the temporal context (as in \cite{Phan2019}, see \ref{section_evaluation}). Finally, our preprocessing is more aggressive than in the original paper. These differences concern only input and output data, not the models themselves. This ensure that all the models are compared in the same conditions of input, preprocessing and prediction.
SimpleSleepNet achieve the best performance on Sleep EDF. On MASS, DeepSleepNet shows the best Macro-F1 score closely followed by SimpleSeepNet.

%% file: tables/parameters_count.tex
\begin{table}[ht]
\vspace{0.7em}
\begin{center}
\scalebox{1.05}{%
\begin{tabular}{|l c c|} 
 \hline
 Model  & $\#$ parameters & training duration (sec.)\\ 
 \hline
SimpleSleepNet-Small & $\num{3.9e04}$ & 123\\ 
SimpleSleepNet  & $\num{9.3e04}$ & 126\\ 
Chambon et al. \cite{Chambon2018} & \num{1.2e5} & 367\\ 
Mixed NN \cite{Dong2018} & \num{2.0e05} & 9.98\\ 
SimpleSleepNet-Large & \num{2.5e05} & 129\\ 
SeqSleepNet \cite{Phan2019} & \num{4.1e05} & 75.8\\ 
DeepSleepNet \cite{Supratak2017}  & \num{1.7e07} & 95.2\\ 
Tsinalis et al. (CNN) \cite{Tsinalis2016a}  & \num{1.7e08} & 268\\ 
 \hline
\end{tabular}
}

\caption{Number of parameters and train time per epoch on a Titan X on the DOD-H dataset. 18 records are used for training and 6 for validation. The order of magnitude and the ranking is the same on DOD-O.}
\label{tab:parameters_count}
\end{center}
\vspace{-4em}
\end{table}

%% file: tables/complete_results.tex
\begin{table*}[ht]
\centering
\scalebox{0.925}{%
\begin{tabular}{|c|c |c c c|c c c c c|} 
 \hline
 & & \multicolumn{3}{|c|}{Overall metrics} & \multicolumn{5}{|c|}{Class-wise F1 (\%)} \\
    &  Model & F1 (\%) & Accuracy (\%)  & $\kappa$ (\%) & W & N1 &N2 &N3 &REM\\ 
 \hline \multirow{10}*{\rotatebox{90}{DOD-H}} & \textbf{SimpleSleepNet}  & \textbf{89.9  $\pm$  4.1}  & \textbf{89.9  $\pm$  4.2} & \textbf{84.6  $\pm$  6.5} &86.1  $\pm$  11.5 &  \textbf{59.8  $\pm$  14.4} &  \textbf{92.4  $\pm$  3.1}  & \textbf{82.5  $\pm$  23.0 }& 91.4  $\pm$  8.3\\ 
 & DeepSleepNet \cite{Supratak2017}  & 89.8 $\pm$ 4.4 & 89.6 $\pm$ 4.4    & 84.3 $\pm$ 6.7 &\textbf{87.1  $\pm$  8.2}&59.2  $\pm$  13.1 &91.9  $\pm$  3.9 &81.2  $\pm$  22.8 &90.0  $\pm$  9.0 \\
 & \textcolor{NavyBlue}{\textit{Scorer (best)}}  &   \textit{\textcolor{NavyBlue}{88.7 $\pm$ 4.7}} & \textit{\textcolor{NavyBlue}{88.7 $\pm$ 4.3}} & \textit{\textcolor{NavyBlue}{82.8 $\pm$ 7.2}}  & \textit{\textcolor{NavyBlue}{85.6  $\pm$  12.0}} & \textit{\textcolor{NavyBlue}{54.7  $\pm$  11.8}} & \textit{\textcolor{NavyBlue}{91.1  $\pm$  3.7}}&\textit{\textcolor{NavyBlue}{78.9  $\pm$  24.8}} & \textit{\textcolor{NavyBlue}{\textbf{91.8  $\pm$  8.0}}}\\
  & SeqSleepNet \cite{Phan2019} & 87.1 $\pm$ 7.7 & 87.2 $\pm$ 7.9  & 80.4 $\pm$ 11.7 &84.6  $\pm$  16.1 &56.6  $\pm$  13.6&89.5  $\pm$  7.1&75.7  $\pm$  25.1& 88.1  $\pm$  11.2\\
  & \textcolor{NavyBlue}{\textit{Scorers (avg.)}} & \textcolor{NavyBlue}{\textit{86.8 $\pm$ 7.6 }} & \textcolor{NavyBlue}{\textit{86.5 $\pm$ 8.1 }}& \textcolor{NavyBlue}{\textit{79.9 $\pm$ 11.3 }}& \textcolor{NavyBlue}{\textit{84.3  $\pm$  13.6 }}& \textcolor{NavyBlue}{\textit{50.3  $\pm$  14.7 }}&\textcolor{NavyBlue}{\textit{88.9  $\pm$  7.6 }}&\textcolor{NavyBlue}{\textit{78.5  $\pm$  23.9}}  & \textcolor{NavyBlue}{\textit{90.8  $\pm$  10.3}} \\
 & Mixed NN \cite{Dong2018} & 86.0 $\pm$ 7.1  & 86.3 $\pm$ 6.3 & 79.1 $\pm$ 9.9 &78.6  $\pm$  15.8&55.9  $\pm$  13.6&89.9  $\pm$  4.8&80.8  $\pm$  22.3 &84.8  $\pm$  18.8\\
 
 & Chambon et al. \cite{Chambon2018}  & 83.0 $\pm$ 4.5  & 83.5 $\pm$ 4.7& 74.6 $\pm$ 7.3 &72.4  $\pm$  13.7&35.5  $\pm$  11.4&87.7  $\pm$  4.1&77.4  $\pm$  22.5& 84.0  $\pm$  11.5\\ 
   & \textcolor{NavyBlue}{\textit{Scorer (worst)}}  & \textcolor{NavyBlue}{\textit{82.4  $\pm$  8.0}} & \textcolor{NavyBlue}{\textit{81.7  $\pm$   8.6}}  & \textcolor{NavyBlue}{\textit{73.3  $\pm$  12.9}} & \textcolor{NavyBlue}{\textit{75.2  $\pm$  17.9}} & \textcolor{NavyBlue}{\textit{40.3  $\pm$  16.5}} &\textcolor{NavyBlue}{\textit{84.7  $\pm$  6.9}} &\textcolor{NavyBlue}{\textit{76.8  $\pm$  21.8}}&\textcolor{NavyBlue}{\textit{91.6  $\pm$  8.9}}\\
 & Tsinalis et al. (CNN) \cite{Tsinalis2016a} & 75.1 $\pm$ 11.7  & 75.7 $\pm$ 11.6 & 61.3 $\pm$ 17.2 &44.0  $\pm$  18.6&15.4  $\pm$  11.5&78.5  $\pm$  12.3 &68.7  $\pm$  24.7& 69.8  $\pm$  18.4  \\ \hhline{|=|=|===|=====|}
 \multirow{10}*{\rotatebox{90}{DOD-O}}  &  \textbf{SimpleSleepNet}  & \textbf{88.3  $\pm$  9.0}  & \textbf{88.7  $\pm$  8.2}  & \textbf{82.3  $\pm$  11.2}  &91.7  $\pm$  7.4&\textbf{55.4  $\pm$  16.8}&\textbf{89.7  $\pm$  10.5}& 64.8  $\pm$  36.0& \textbf{86.5  $\pm$  22.5}\\ 

 & \textcolor{NavyBlue}{\textit{Scorers (best)}} & \textcolor{NavyBlue}{\textit{88.0  $\pm$  7.0}}  & \textcolor{NavyBlue}{\textit{87.5  $\pm$  6.7}}  & \textcolor{NavyBlue}{\textit{80.5  $\pm$  9.5}} &\textcolor{NavyBlue}{\textit{\textbf{92.7  $\pm$  6.9}}} &\textcolor{NavyBlue}{\textit{48.5  $\pm$  15.3}}&\textcolor{NavyBlue}{\textit{88.3  $\pm$  8.5}}&\textcolor{NavyBlue}{\textit{63.7  $\pm$  33.9}}&\textcolor{NavyBlue}{\textit{\textbf{86.5  $\pm$  22.4}}}\\
  & DeepSleepNet \cite{Supratak2017}  & 87.5  $\pm$  8.5  & 87.5  $\pm$  8.6  & 80.4  $\pm$  12.2 &88.6  $\pm$  10.1 &49.2  $\pm$  17.2&89.1  $\pm$  9.1 &\textbf{68.0  $\pm$  31.4} &85.5  $\pm$  21.6 \\
    & SeqSleepNet \cite{Phan2019} & 85.1  $\pm$  10.3 & 85.5  $\pm$  10.1  & 77.2  $\pm$  14.8 & 86.6  $\pm$  12.6 &48.6  $\pm$  21.6 & 88.1  $\pm$  10.2& 60.8  $\pm$  34.4 & 79.4  $\pm$  28.0\\
       & Chambon et al. \cite{Chambon2018} & 85.0  $\pm$ 8.3 & 85.1  $\pm$  8.3 & 76.6  $\pm$  12.0 &84.0  $\pm$  12.3&41.3  $\pm$  17.0&88.0  $\pm$  7.9 &60.6  $\pm$  35.1&81.4  $\pm$  22.2 \\ 
  & \textcolor{NavyBlue}{\textit{Scorers (avg.)}}& \textcolor{NavyBlue}{\textit{84.8  $\pm$  8.6}} & \textcolor{NavyBlue}{\textit{85.0  $\pm$  8.9}} & \textcolor{NavyBlue}{\textit{76.5  $\pm$  12.3}} &\textcolor{NavyBlue}{\textit{90.8  $\pm$  8.2}}  &\textcolor{NavyBlue}{\textit{44.9  $\pm$  16.2}}& \textcolor{NavyBlue}{\textit{85.6  $\pm$  10.7}} &\textcolor{NavyBlue}{\textit{56.9  $\pm$  33.1}} &\textcolor{NavyBlue}{\textit{85.6  $\pm$  23.3}}  \\

  & \textcolor{NavyBlue}{\textit{Scorer (worst)}}  & \textcolor{NavyBlue}{\textit{82.8  $\pm$  10.0}} & \textcolor{NavyBlue}{\textit{84.3  $\pm$   8.3}} & \textcolor{NavyBlue}{\textit{75.4  $\pm$  12.4}} &\textcolor{NavyBlue}{\textit{90.9  $\pm$  7.2}}&\textcolor{NavyBlue}{\textit{84.5  $\pm$  12.0}} &\textcolor{NavyBlue}{\textit{44.5  $\pm$  16.5}} &\textcolor{NavyBlue}{\textit{46.6  $\pm$  33.3}}&\textcolor{NavyBlue}{\textit{\textbf{86.5  $\pm$  21.8}}}\\
  
   & Mixed NN \cite{Dong2018}  & 81.9  $\pm$  11.2 & 82.1  $\pm$  11.2 & 72.5  $\pm$  15.2 &84.8  $\pm$  13.9 &49.7  $\pm$  16.1&85.6  $\pm$  9.7  &61.8  $\pm$  33.5  &71.8  $\pm$  27.1 \\

 & Tsinalis et al. (CNN) \cite{Tsinalis2016a}& 79.1  $\pm$  10.4  & 79.4  $\pm$  9.6   & 67.6  $\pm$  14.0 &76.2  $\pm$  19.8 &30.6  $\pm$  14.4 &84.0  $\pm$  9.9 &59.8  $\pm$  30.3 &66.3  $\pm$  26.3 \\ 
 \hline
\end{tabular}
}
\caption{Performance metrics of each of the baseline models. Average, best and worst human scorers performance are also given. The best (resp. worse) scorer is the scorer with the highest (lowest) F1.}
\label{tab:complete_results}
\vspace{-1.6em}
\end{table*}

%% file: tables/ablations.tex
\begin{table}[ht]
\centering
\scalebox{0.90}{%
\begin{tabular}{|c |c c c c|} 
 \hline
 
 & Model & F1 (\%) & Accuracy (\%) & $\kappa$ (\%) \\ 
 \hline
 \multirow{6}*{\rotatebox{90}{DOD-H}} & \textbf{No channel recombination} & \textbf{90.2 $\pm$ 4.2} & \textbf{90.2 $\pm$ 4.2}  & \textbf{85.0 $\pm$  6.5} \\
 & No attention & 90.1 $\pm$ 4.2 & 90.1 $\pm$  4.1   &  \textbf{85.0 $\pm$ 6.1} \\
 & \textit{SimpleSleepNet}  & \textit{89.9 $\pm$ 4.1} & \textit{89.9 $\pm$ 4.2 }  & \textit{84.6 $\pm$ 6.5 }  \\
  & No skip connections & 89.9 $\pm$ 3.9 & 89.9 $\pm$ 3.8    & 84.6 $\pm$ 6.1  \\
 & No positional embeddings & 89.8 $\pm$ 3.2 & 89.9 $\pm$ 3.2    & 84.6 $\pm$ 4.9  \\
 & Single GRU layer & 89.6 $\pm$ 4.1  & 89.6 $\pm$ 4.1  & 84.1 $\pm$ 6.2  \\
  & No filtering & 89.2 $\pm$ 4.6 & 89.3 $\pm$ 4.5    & 83.6 $\pm$ 6.8  \\
  & Filter bank & 89.1 $\pm$ 4.1 & 89.0 $\pm$ 4.3    & 83.3 $\pm$ 6.7  \\
  & No channel dropout & 87.9 $\pm$ 8.0 & 88.4 $\pm$ 6.7    & 82.2 $\pm$ 10.4  \\
 & No frequency reductions & 87.5 $\pm$ 7.6 & 87.9 $\pm$ 7.1   & 81.5 $\pm$ 10.5  \\
 \hhline{|=|====|}
  \multirow{6}*{\rotatebox{90}{DOD-O}} & \textbf{ SimpleSleepNet}& \textbf{88.3 $\pm$ 9.0} & \textbf{88.7 $\pm$ 8.2}   & \textbf{82.3 $\pm$ 11.2}  \\
 & \textbf{Single GRU layer} & \textbf{88.3 $\pm$ 7.3} & 88.4 $\pm$ 7.3    & 81.9 $\pm$ 10.5  \\
 & No skip connections  & 88.2 $\pm$ 9.3  & 88.3 $\pm$ 8.9  & 81.8 $\pm$ 12.7  \\
  & No positional embeddings & 88.1 $\pm$ 7.9  & 88.4 $\pm$ 7.7    & 81.8 $\pm$ 11.0  \\
 & No Attention & 87.6 $\pm$ 10.5 & 87.9 $\pm$ 10.0  & 81.2 $\pm$ 13.3\\
 & No channel recombination  & 87.5 $\pm$ 6.7 & 87.9 $\pm$ 6.2 & 80.8 $\pm$  9.6 \\
 & Filter bank & 86.9 $\pm$ 8.8 & 87.1 $\pm$ 8.3    & 79.6 $\pm$ 13.0  \\
 & No filtering   & 86.2 $\pm$ 13.1  & 86.4 $\pm$ 12.2& 79.4 $\pm$ 15.1 \\
  & No channel dropout   & 85.3 $\pm$ 13.3  & 85.6 $\pm$ 11.9& 77.9 $\pm$ 15.7 \\
 & No frequencies reduction & 85.0 $\pm$ 9.7  & 85.4 $\pm$ 9.7 & 77.2 $\pm$ 14.0 \\
 \hline
\end{tabular}
}
\caption{Performance metrics of ablated variations of SimpleSleepNet. For each model, a specific module from SimpleSleepNet is either removed or replaced by a simpler alternative.}
\label{tab:ablations}
\vspace{-3.3em}
\end{table}

%% file: tables/model_size.tex
\begin{table}[ht]
\centering
\scalebox{0.94}{%
\begin{tabular}{|c|c c c c|} 
 \hline
 & Model & F1 (\%) & Accuracy (\%) & $\kappa$ (\%) \\ 
 \hline
  \multirow{3}*{\rotatebox{90}{\scriptsize{DOD-H}}}   & \textbf{SimpleSleepNet-Large}  & \textbf{90.4 $\pm$ 3.8 } & \textbf{90.4 $\pm$ 3.8} & \textbf{85.3 $\pm$ 6.0 }  \\ 
 & SimpleSleepNet   & 89.9 $\pm$ 4.1  & 89.9 $\pm$ 4.2 & 84.6 $\pm$ 6.5 \\ 
   & SimpleSleepNet-Small & 89.3 $\pm$ 4.6  & 89.2 $\pm$ 4.5 & 83.7 $\pm$ 6.9 \\ 
     \hhline{|=|====|}\multirow{3}*{\rotatebox{90}{\scriptsize{DOD-O}}}& \textbf{SimpleSleepNet-Large} & \textbf{89.0 $\pm$ 6.0} & \textbf{89.0 $\pm$ 6.6}   & \textbf{83.0 $\pm$ 9.1 } \\ 
 & SimpleSleepNet & 88.3 $\pm$ 9.0   & 88.7 $\pm$ 8.2  & 82.3 $\pm$ 11.2  \\ 
    & SimpleSleepNet-Small & 87.2 $\pm$ 10.4 & 87.7 $\pm$ 9.3  & 80.9 $\pm$ 12.3  \\
   
 \hline
\end{tabular}
}
\caption{Performance metrics of SimpleSleepNet variants with smaller (SimpleSleepNet-Small) and larger (SimpleSleepNet-Large) layer size than the original models. }
\label{tab:model_size}
\vspace{-2em}
\end{table}

%% file: tables/single_channel.tex
\begin{table}[ht]
\centering
\begin{tabular}{|c|c c c c|} 
 \hline
 & Model  & F1 (\%)  & Accuracy (\%)& $\kappa$ (\%) \\ 
 \hline
 \multirow{3}*{\rotatebox{90}{\scriptsize{DOD-H}}}  & \textbf{All channels} & \textbf{89.9 $\pm$ 4.2}  & \textbf{89.9 $\pm$ 4.1 }  & \textbf{84.6 $\pm$ 6.5 }  \\
    & \textcolor{NavyBlue}{\textit{Scorers (avg.)}} & \textcolor{NavyBlue}{\textit{86.8 $\pm$ 7.6}} & \textcolor{NavyBlue}{\textit{86.5 $\pm$ 8.1}}  & \textcolor{NavyBlue}{\textit{79.9 $\pm$ 11.3}}  \\
  & F4-O2 & 86.6 $\pm$ 7.7 & 86.7 $\pm$ 7.4  & 79.7 $\pm$ 11.3  \\
   \hhline{|=|====|} \multirow{3}*{\rotatebox{90}{\scriptsize{DOD-O}}} & \textbf{All channels} & \textbf{88.3 $\pm$ 9.0} & \textbf{88.7 $\pm$ 8.2 }   & \textbf{82.3 $\pm$ 11.2 }  \\
    & \textcolor{NavyBlue}{\textit{Scorers (avg.)}}& \textcolor{NavyBlue}{\textit{84.8 $\pm$ 8.6}} &  \textcolor{NavyBlue}{\textit{85.0 $\pm$ 8.9}} & \textcolor{NavyBlue}{\textit{76.5 $\pm$ 12.3}} \\
 &  F4-O2 & 84.4 $\pm$ 8.5 & 84.4 $\pm$ 9.3  & 75.8 $\pm$ 14.5  \\
 
 \hline
\end{tabular}
\caption{Performance metrics are compared when SimpleSleepNet is trained on the F4-02 derivation only vs when it is trained on all PSG channels. The Scorers (avg.) from Table \ref{tab:complete_results} is given for reference.}
\label{tab:single_channel}
\vspace{-2em}
\end{table}

%% file: tables/transfer_learning.tex
\begin{table}[ht]
\vspace{1em}
\centering
\begin{tabular}{|c c c c|} 
 \hline
 Model & F1 (\%) & Accuracy (\%)  & $\kappa$ (\%) \\ 
  \hline
     DOD-H (scratch) & 89.9 $\pm$ 4.1& 89.9 $\pm$ 4.2  & 84.6 $\pm$ 6.5 \\ 
      DOD-O to DOD-H & 84.8 $\pm$ 7.8  & 84.9 $\pm$ 8.1   & 77.6 $\pm$ 11.5  \\
\hhline{|====|}
 DOD-O (scratch)& 88.3 $\pm$ 9.0 & 88.7 $\pm$ 8.2  & 82.3 $\pm$ 11.2  \\ 
    DOD-H to DOD-O & 62.6 $\pm$ 14.9& 68.3 $\pm$ 12.4  & 45.7 $\pm$ 17.2  \\
    \hline
\end{tabular}
\caption{Performance metrics of SimpleSleepNet when trained on DOD-O (resp. DOD-H) and evaluated on the other dataset DOD-H (resp. DOD-O). The models are trained and validated on 20 random splits of the source dataset and evaluated against the target dataset.}
\label{tab:transfer_learning}
\vspace{-2em}
\end{table}

%% file: tables/external_datasets.tex
\begin{table*}[ht]
\centering
\scalebox{1.15}{%
\begin{tabular}{|c|c|c|c|c|c|} 

 \hline
Model  & \multicolumn{2}{c}{MASS}  & \multicolumn{2}{c}{S-EDF-20} & S-EDF-Extended  \\
 &  MF1 (\%) (Ours) & MF1 (\%) (Theirs) &  MF1 (\%) (Ours) & MF1 (\%) (Theirs) & MF1 (\%) (Ours) \\
 \hline
SimpleSleepNet & 84.7  & - & \textbf{80.5} & -  & \textbf{79.1}\\
DeepSleepNet \cite{Supratak2017} & \textbf{85.2} & 81.7 & \textit{80.4} & 76.9  & 78.1\\

SeqSleepNet \cite{Phan2019}& 83.2 &83.3 \textsuperscript{(*)}  & 80.0 & - & \textit{78.5} \\
Mixed Neural Network  \cite{Dong2018} & 83.2 \textsuperscript{(1)} & 80.5  &  75.8 \textsuperscript{(2)}& - & 75.6 \textsuperscript{(2)}\\

Chambon et Al.  \cite{Chambon2018} & 79.0 & 76.7 & 75.9 & - & 76.6 \\
Tsinalis et Al. \cite{Tsinalis2016} & 77.1 & 70.4 \textsuperscript{(*)} & 72.2 &  69.8 & 74.7  \\
U-Time \cite{utime} & -  & - & - & 79.0  & -\\
 
 \hline
\end{tabular}
}
\caption{Macro-F1 of the baseline models on MASS and Sleep EDF. For consistency with the literature we report the epoch-wise Macro F1. Moreover, since the computation are done epoch-wise, we cannot report subject variability as in the other tables. (*) are reported on the complete MASS dataset.\textsuperscript{(1)} is trained on F4-EOG and \textsuperscript{(2)} on Fpz-Cz to limit the number of parameters.  }
\label{tab:mass_evaluation}
\vspace{-2.6em}
\end{table*}

%% file: 4_discussion.tex
\section{Discussion}

DOD-H and DOD-O multiple scoring highlight the previously described and relatively high inter-rater variability regarding sleep staging. This confirms the need for automated sleep staging approaches to train and compare with a consensus of human scorers instead of a single human scorer for a more realistic evaluation of performance. The $\varB{Soft-Agreement}$ and the methodology presented allow to handle multiple scorers and especially situations when a tie between scorers occurs. Another solution could be using yet more scorers to reduce ties occurrence and improve the fairness of the built consensus.

Due to an increased sleep fragmentation, manual sleep staging is more difficult on patients with OSA than healthy subjects. This is also true for most automated approaches. Indeed, the accuracy is lower and presents higher variance on DOD-O than on DOD-H. There are also more ties on DOD-O than DOD-H. This is in agreement with \cite{Stephansen2018} where models accuracy drops by 9 \% on narcoleptic subjects vs healthy subjects and with \cite{penzel2013} where the scorers reliability was much higher on healthy subjects than on those with OSA. Besides, the training requires more recordings to reach human performance on DOD-O than on DOD-H. All those elements suggest that the inter-subjects variability is higher within DOD-O than within DOD-H. 
Yet, interestingly, transfer learning from DOD-O to DOD-H is much more effective than the other way around. This implies that data acquired from patients suffering from OSA contains information related to healthy sleep as well as information specific to OSA. This also shows that although SimpleSleepNet reaches a better F1 on DOD-H than on DOD-O, the model trained on DOD-O is much better in its generalization capacity than the one trained on DOD-H. These analyses could be extended to datasets with other sleep-related issues to see how much they impact the performance of human and automated sleep staging. This also suggests that a dataset containing high inter-subject variability, for instance with a mix of both abnormal and normal sleep, would probably lead to better models in terms of their ability to generalize. This is also highlighted in \cite{Stephansen2018}.

The transfer learning experiment also highlights a practical limitation regarding the usability of such automatic method outside of the scope of the same population of patients or/and device than the one on which it has been trained on. This is also discussed in \cite{Chambon2018transfer}. In practice, this limits the use of such automatic sleep staging method in a clinical setup. Training models on a cohort of several patients with a mix of both abnormal and normal sleep recorded on different PSG devices and scored by different scorers would greatly improve the generalization of sleep staging models. However, the use of different devices implies dealing with possibly different modalities and missing signals, which is a problem that has to our knowledge not been tackled yet.

SimpleSleepNet, DeepSleepNet and SeqSleepNet outperform the average human scorer on both DOD-O and DOD-H. Most other automated approaches perform with an accuracy close to human scorers. The confusion matrix also shows similar pattern of mistakes between humans and SimpleSleepNet. Given a few annotated records, automated sleep staging could reach similar performances to human scorers in a clinical setting if the data are acquired with a consistent PSG montage and patient typology. This is often the case in a typical sleep clinic setting. That being said, an interesting direction of research would be to create a model able to adapt to various PSG montage without fine-tuning or weight modifications.

On external datasets, SimpleSleepNet, DeepSleepNet, and SeqSleepNet also show the best performances. However, these datasets were scored by a single expert. Inter-rater variability prevents us from drawing strong conclusions regarding the absolute performance of the various models on these datasets. Specifically, the models could be overfitting on human expert scoring.

We observe that most benchmarked methods using data-driven feature extraction perform better than the expert feature extraction approach. This is especially true on DOD-O and SleepEDF-Extended which present a higher level of variability, suggesting a better ability for such deep learning models to capture relevant information in complex data like abnormal sleep.

SimpleSleepNet outperforms the best human scorer and all other sleep staging models on DOD-O and DOD-H. It is also among the best-ranked models on external datasets. It uses significantly fewer parameters than other approaches. The presented ablation study shows that the various building blocks of SimpleSleepNet allow reaching the best performance on DOD-O. SimpleSleepNet reaches close-to-human performance with only a few ($\sim$10) recordings, suggesting that sleep stage classification is a relatively simple problem in terms of data quantity needed to reach satisfactory performance. The temporal context and number of signals also seem to play a minor role in improving performance.

The results provided in this study are available with both data and code for reproducibility. It should be noted that the benchmarked automated approaches were all reimplemented. The performances of our implementation were validated on the MASS SS3 and SleepEDF datasets with performance similar or above the original implementations.
Furthermore, to ensure the fairness of the benchmark, every method was tuned to provide good results on the datasets of this study. All reported results are from a single run, rerunning the experiments might result in slightly different results due to randomness and variability.

%% file: 5_conclusion.tex
\section{Conclusion}
In this work, we introduced two open multi-scored sleep staging datasets with 25 from healthy subjects and 55 nights patients suffering from OSA. We proposed a methodology for evaluation against multiple human scorers. We showed the relevance of a multi-scored sleep dataset to assess how automated sleep staging performs in a clinical setting. We demonstrated that recent automated sleep staging performances are often on-par with the average human scorer, and that the best automated sleep staging are better than the best human scorer. We also introduced a new efficient sleep staging model, SimpleSleepNet, which outperforms previous state-of-the-art models and human scorers on both datasets and on two frequently benchmarked datasets. Better understanding and quantification of the performance of such automated approaches could be a step toward a broader use of these approaches in sleep clinics.